\documentclass[aps,prx,reprint,twocolumn,floatfix,superscriptaddress,amsmath,amssymb,amsfonts,nofootinbib,balancelastpage]{revtex4-2}

\allowdisplaybreaks
\usepackage{comment}
\usepackage{xcolor}
\usepackage{url}
\usepackage{graphicx}
\usepackage{amsmath,amssymb,amstext,amsfonts,bm,dsfont}
\usepackage{siunitx}
\usepackage{enumitem}
\usepackage{hyperref}
\hypersetup{
    colorlinks=true,
    linkcolor=blue,
    citecolor=blue,
    filecolor=blue,
    urlcolor=blue,
    breaklinks=true,
    unicode=true,
    pdfauthor={Andreas Feuerpfeil, Ronny Thomale, Trithep Devakul, Nicolas Regnault, Julian May-Mann},
    pdfsubject={Condensed Matter Physics},
    pdfdisplaydoctitle=true
}

\usepackage{tabularray}
\UseTblrLibrary{booktabs}
\usepackage{slashed,physics}
\usepackage{orcidlink}

\usepackage{zref-clever}
\zcsetup{cap,abbrev,nameinlink=false,comp}

\zcRefTypeSetup{equation}{name-sg = equation, name-pl = equations, name-sg-ab = eq., name-pl-ab = eqs.}
\zcRefTypeSetup{figure}{name-sg = figure, name-pl = figures, name-sg-ab = fig., name-pl-ab = figs.}
\zcRefTypeSetup{appendix}{name-sg = appendix, name-pl = appendices, name-sg-ab = app., name-pl-ab = apps.}
\zcRefTypeSetup{section}{name-sg = section, name-pl = sections, name-sg-ab = sec., name-pl-ab = secs.}
\zcRefTypeSetup{table}{name-sg = table, name-pl = tables, name-sg-ab = tab., name-pl-ab = tabs.}

\newcommand{\hiddenlink}[2]{\hyperref[#1]{#2}}
\NewDocumentCommand{\figref}{ o m }{%
    \IfNoValueTF{#1}%
        {\zcref{#2}}%
        {\hiddenlink{#2}{\zcref*{#2}(#1)}}%
}

\newcommand{\mbf}[1]{\mathbf{#1}}

\NewDocumentCommand{\intk}{ O{d} }{ \int \frac{\dd^{#1}k}{(2\pi)^{#1}} }
\NewDocumentCommand{\intp}{ O{d} }{ \int \frac{\dd^{#1}p}{(2\pi)^{#1}} }
\NewDocumentCommand{\intq}{ O{d} }{ \int \frac{\dd^{#1}q}{(2\pi)^{#1}} }

\begin{document}

\title{Nature of the bosonic edge phase in a double quantum spin Hall insulator}
\author{Andreas Feuerpfeil}
\email{andreas.feuerpfeil@uni-wuerzburg.de}
\affiliation{Institut für Theoretische Physik und Astrophysik, Julius-Maximilians-Universität Würzburg, Am Hubland, Campus Süd, 97074 Würzburg, Germany}
\affiliation{Würzburg–Dresden Cluster of Excellence ctd.qmat, 97074 Würzburg and 01062 Dresden, Germany}
\affiliation{Google, Mountain View, California 94043, USA}
\author{Ronny Thomale}
\affiliation{Institut für Theoretische Physik und Astrophysik, Julius-Maximilians-Universität Würzburg, Am Hubland, Campus Süd, 97074 Würzburg, Germany}
\affiliation{Würzburg–Dresden Cluster of Excellence ctd.qmat, 97074 Würzburg and 01062 Dresden, Germany}
\author{Trithep Devakul}
\affiliation{Department of Physics, Stanford University, Stanford, CA 94305, USA}
\author{Nicolas Regnault}
\affiliation{Center for Computational Quantum Physics, Flatiron Institute, 162 5th Avenue, New York, NY 10010, USA}
\affiliation{Laboratoire de Physique de l'Ecole normale sup\'{e}rieure, ENS, Universit\'{e} PSL, CNRS, Sorbonne Universit\'{e}, Universit\'{e} Paris-Diderot, Sorbonne Paris Cit\'{e}, 75005 Paris, France}
\author{Julian May-Mann}
\email{maymann@stanford.edu}
\affiliation{Department of Physics, Stanford University, Stanford, CA 94305, USA}

\begin{abstract}
Field-theoretic analysis has predicted that the edge of a double quantum spin Hall insulator (DQSHI) can undergo an interaction-driven transition from a phase with fermionic edge modes into one with only bosonic modes.
All single-fermion edge excitations are gapped in this bosonic edge phase, but fermion pairs remain gapless, leading to distinct experimental signatures. We numerically simulate the interacting DQSHI edge and verify the existence of the bosonic edge phase. This calculation requires an explicit bulk embedding of the DQSHI edge, due to its mixed anomaly between charge and $S^z$ conservation. We hence employ segment DMRG on an infinite two-dimensional cylinder, where the DQSHI bulk is modeled in terms of two copies of the spin-up lowest Landau level, combined with their time-reversed counterparts. We find that interactions can open a single-fermion gap at the DQSHI edge while fermion pairs remain gapless, matching the predictions for the bosonic edge phase. We find the scaling of the single-fermion gap with interaction strength to be consistent with field-theoretic calculations, and  discuss the relevance of the bosonic edge phase to twisted homobilayers of WSe$_2$ and MoTe$_2$. We  elaborate on the methodological extension of the segment geometry used here to other topological phases with edge anomalies.
\end{abstract}

\maketitle

\section{Introduction}
Topological phases of matter characteristically display a bulk-boundary correspondence, where the gapped topological bulk is accompanied by gapless edge modes~\cite{hasan2010colloquium,qi2011topological, mong2011edge, essin2011bulk, rhim2018unified}. However, this is not a one-to-one relationship. The same bulk can realize multiple different edges, and the same edge can be realized by different bulk phases~\cite{cano2014bulk}. While there are canonical instances of such ambiguity, as it is always possible to attach a lower-dimensional subsystem to a given edge~\cite{ChamonWen1994sharp, Yang2003field, Amaricci2017edge}, there are also  systems that host multiple edges with fundamentally distinct physical properties emanating from the same topological bulk theory~\cite{haldane1995stability, plamadeala2013short, barkeshli2015superconductivity, lichtman2021bulk}.

\begin{figure}
    \centering
    \includegraphics[width=\linewidth]{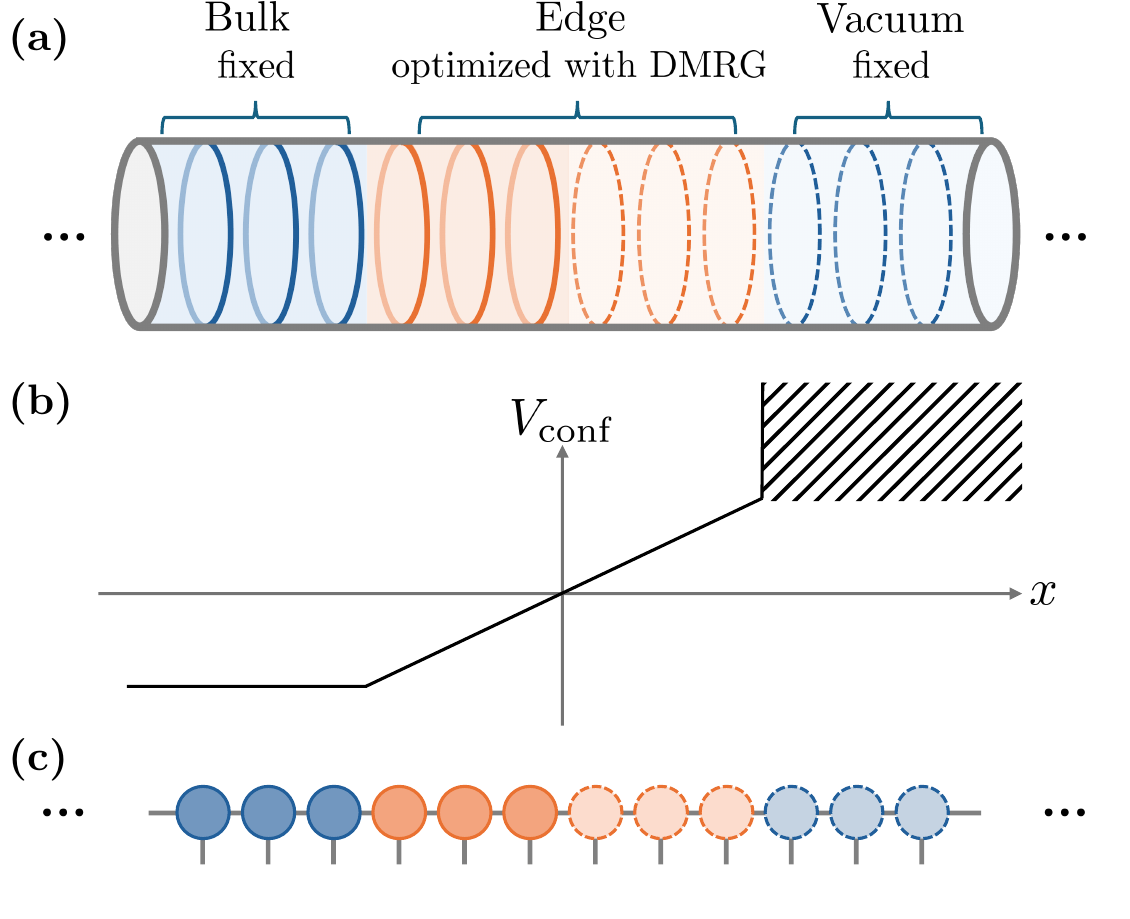}
    \caption{\textbf{Topological edge mode analysis in DMRG.} For a single chiral edge mode, the lowest Landau level wave functions are defined on a cylinder \textbf{(a)} subject to a confining potential \textbf{(b)}. Without interactions, the states on the left are filled (solid rings), the states on the right are empty (dashed rings), and states at the edge have a chiral dispersion. The DQSHI consists of two copies of this cylinder and their time-reversed counterparts. To analyze the effects of interactions at the edge, the cylinder is expressed as a matrix product state \textbf{(c)}. DMRG is then performed on the finite middle segment (orange) only, with the exterior segments (blue) held fixed.}
    \label{fig:segment_geo}
\end{figure}

One of the paradigmatic systems where this occurs is the double quantum spin Hall insulator (DQSHI)~\cite{ezawa2013high, bai2022doubled, liu2024orbital, xue2023antiferromagnetic, hung2025tunable, ye2026double}. The DQSHI is a topological phase protected by charge conservation and $S^z$-spin conservation, with spin Chern number $C_s = 2$~\cite{sheng2006quantum}. Since the DQSHI bulk is adiabatically equivalent to two decoupled (single) quantum spin Hall insulators (QSHIs)~\cite{KaneMele_2005a,KaneMele_2005b,BernevigZhang_2006,BHZ_2006,Konig_2007,MaciejkoHughesZhang_2011}, one would expect the DQSHI edge to host two pairs of helical fermionic modes, i.e., two chiral spin-up fermions and two anti-chiral spin-down fermions. However, continuum field theory analysis has shown that this is not always the case. Under consideration of electronic interactions, the DQSHI edge can undergo a transition to a new bosonic edge phase with the very same symmetries yet only one pair of non-chiral bosonic edge modes~\cite{bi2017bilayer, may2025theory}. All single-fermion excitations are gapped in this phase while fermion pairs, which can be recast as bosons, are gapless. Importantly, the fermion-to-boson edge transition can occur without gap closing in the DQSHI bulk~\cite{Tsantilas_2026}.

Experimentally, the fermionic and bosonic edge phases can be distinguished in shot-noise measurements using a quantum point contact geometry~\cite{KaneFisher_1994,Saminadayar_1997,dePicciotto_1997}.
In the fermionic edge phase, charge is backscattered in units of $e$, leading to a Fano factor of $F = 1$. In the bosonic edge phase, however, the single-fermion gap only allows for backscattering in units of $2e$, producing a doubled Fano factor of $F=2$~\cite{zhang2017fingerprints, Tsantilas_2026}. The DQSHIs recently observed in twisted homobilayers of the transition metal dichalcogenides
(TMDs) MoTe$_2$~\cite{Kang_2024_MoTe2,Kang_2025_TRSB,xu2025interplay} and WSe$_2$~\cite{Kang_2024_WSe2, Han2025QuantumOscillations}  are natural candidates for such shot-noise measurements, as the quantum point contact geometry can be directly integrated into the device architecture~\cite{Ren_2026_FerroelectricQPC}.

While the bosonic edge phase of the DQSHI is well established within the framework of continuum field theory, it has yet to be confirmed that this phase can be realized in an explicit microscopic model. This requires a numerical approach. One may expect that the density-matrix renormalization group (DMRG) would be well suited for this task, as it is an exceptionally powerful tool for studying strongly correlated 1D systems~\cite{white1992density, white1993density, schollwock2005density,hastings2007area, schollwock2011density}. Still, the edge modes of the DQSHI are \textit{anomalous}~\cite{chen2011two, ryu2012electromagnetic, wen2013classifying, kapustin2014anomalies, else2014classifying}: specifically, they carry a mixed anomaly between spin- and charge-conservation, which is canceled out by the topological bulk of the DQSHI~\cite{callan1985anomalies, wen1991gapless, maeda1996chiral,cho2011topological}. This anomaly is an obstruction to realizing the DQSHI edge modes in a symmetric and local 1D lattice model. Simulating the DQSHI edge in DMRG in a strictly 1D geometry would therefore require either breaking a relevant symmetry of the edge or working in a momentum-space basis.
The former is inappropriate for studying a symmetry-protected topological phase. The latter leads to volume-law entanglement entropy and to an anti-sparse, i.e., all-to-all, Hamiltonian upon inclusion of interactions, rendering it impossible to scale up the calculations to larger system sizes.

We circumvent these obstacles by using segment DMRG~\cite{phien2012infinite, mong2017fibonacci, zhu2020topological, divic2025anyon}, which allows for a generic 2D edge embedding. The starting point for this is an infinite 2D cylinder with an interface between the vacuum and the topological DQSHI bulk as shown in Fig.~\ref{fig:segment_geo}. The latter is modeled in terms of four copies of the lowest Landau level (LLL): two spin-up with positive chirality and two spin-down with negative chirality. The edge modes naturally arise at the vacuum interface~\cite{halperin1982quantized, macdonald1984quantized, tong2016lectures}. In particular, this setup is anomaly free, as both the bulk and edge are present. To proceed, we partition the cylinder into a semi-infinite bulk segment, a semi-infinite vacuum segment, and a finite edge segment. We then perform DMRG on the finite edge segment, keeping the bulk and vacuum segments as fixed boundary conditions. While we are interested in the DQSHI, our construction can be generalized to other topological phases of matter, including those hosting fractionalized degrees of freedom and enriched symmetries~\cite{wen2004quantum,PhysRevB.90.205101}.

Using this setup, we calculate the single-fermion gap and the fermion-pair gaps of the interacting DQSHI edge. In the parameter regimes where bosonization predicts the bosonic edge phase, we indeed find a single-fermion gap that extrapolates to a finite value in the thermodynamic limit. Moreover, the single-fermion gap has the exponential dependence on interaction strength that is predicted by bosonization. The fermion-pair gaps, in contrast, remain consistent with zero for all interaction strengths studied. These
results show that the bosonic edge phase can arise in a microscopic DQSHI model, supporting moiré TMDs as candidate platforms for its realization.

The remainder of this Article is organized as follows. In Section~\ref{sec:bosonicEdgeReview} we review the DQSHI and the bosonization analysis of the edge, followed by Section~\ref{sec:segmentGeo} where we discuss the segment geometry setup and implementation. In Section~\ref{sec:results} we realize the bosonic edge phase of the DQSHI using segment DMRG, and show it to exhibit the correct gap structure. 
We conclude our work in Section~\ref{sec:discussion}, where we discuss the relevance of our DQSHI results to twisted MoTe$_2$ and WSe$_2$ homobilayers.

\section{Bosonic edge phase}\label{sec:bosonicEdgeReview}

We begin by reviewing the fermionic and bosonic edge phases of the DQSHI within the framework of bosonization~\cite{bi2017bilayer,may2025theory, Tsantilas_2026}. Consider a DQSHI formed by filling two spin-up bands with Chern number $+1$ and two spin-down bands with Chern number $-1$. Without interactions, these bands contribute four fermionic edge modes, $\psi_{\tau \sigma}$, which are labeled by their spin $\sigma = \uparrow,\downarrow$ and band index $\tau = 1,2$ (in the limit of two decoupled QSHIs, $\tau$ labels the two copies). Using bosonization, we express the fermionic modes as~\cite{fradkin2013field,gogolin2004bosonization}
\begin{equation}
\psi_{\tau \sigma}\propto e^{i\phi_{\tau \sigma}+ik_{F \tau \sigma}x},
\label{eq:Psi}
\end{equation}
where  $k_{F \tau \sigma}$ is the Fermi momentum. Due to time-reversal symmetry, $k_{F \tau \uparrow} = -k_{F \tau \downarrow}$. Without interactions, the bosonized edge Lagrangian is given by
\begin{equation}
    \mathcal{L}_{\mathrm{edge}} = -\frac{1}{4\pi}\partial_x\bm{\phi}^\intercal\bm{K}\,\partial_t\bm{\phi} + \frac{1}{4\pi}\partial_x\bm{\phi}^\intercal\bm{V}\,\partial_x\bm{\phi},
    \label{eq:L0}
\end{equation}
with $\bm{\phi}=(\phi_{1\uparrow},\phi_{1 \downarrow},\phi_{2 \uparrow},\phi_{2 \downarrow})^\intercal$ and $\bm{K}=\mathrm{diag}(1,-1,1,-1)$. $\bm{V}$ is the velocity matrix, which is diagonal and positive in the absence of interactions. From the equations of motion for the bosons, we find that the $\sigma = \uparrow$ modes are all right-moving and the $\sigma = \downarrow$ modes are all left-moving, as expected for the DQSHI edge. The charge and $S^z$-spin conservation of the DQSHI respectively lead to the symmetries $\phi_{\tau \sigma}\to\phi_{\tau \sigma}+\theta_c$ and $\phi_{\tau \sigma}\to\phi_{\tau \sigma}+\sigma_z \theta_s$, where $\sigma_z = \pm 1$ for $\sigma = \uparrow,\downarrow$. Here and throughout, we will assume that time-reversal symmetry is present, although the DQSHI edge is stable without this symmetry~\cite{may2025theory}.

In general, there are two types of interactions compatible with charge-, spin-, and momentum-conservation: a ``double spin-flip''  interaction and the usual density-density interactions. Within bosonization, the double spin-flip interaction is
\begin{equation}\begin{split}
    :\!\psi^\dagger_{1\uparrow}\psi_{2\downarrow}\,\psi^\dagger_{1\downarrow}\psi_{2\uparrow}\!: \!+ \text{h.c.}
    &\propto \cos(\phi_{1\uparrow}\!+\!\phi_{1\downarrow}\!-\!\phi_{2\uparrow}\!-\!\phi_{2\downarrow}),
\end{split}\label{eq:dsf}\end{equation}
where $:\!\dots\!:$ denotes normal ordering with respect to the non-interacting Fermi liquid. Note that this interaction conserves $S^z$-spin, but mixes different $\tau$-sectors.  
The density-density interactions take the form
\begin{equation}
:\!\psi^\dagger_{\tau \sigma}\psi_{\tau \sigma}\,\psi^\dagger_{\tau'\sigma'}\psi_{\tau'\sigma'}\!: \,\,\,\propto\partial_x\phi_{\tau \sigma}\partial_x\phi_{\tau'\sigma'}.
\label{eq:densitydensity}\end{equation}
Since the density-density interactions are quadratic in $\phi_{\tau\sigma}$, they can be absorbed into the velocity matrix, $\bm{V}$. To account for this, we consider $\bm{V}$ parametrized as
\begin{equation}
    \bm{V} = \begin{pmatrix}
        v & -v' & w & -w'\\
        -v' & v & -w' & w\\
        w & -w' & v & -v'\\
        -w' & w & -v' & v
    \end{pmatrix}.
    \label{eq:V}
\end{equation}
In using this parametrization, we have assumed that the Lagrangian is invariant under exchanging the band index $\tau$. Relaxing this assumption does not change any central results~\cite{Tsantilas_2026}.

To discuss the effects of these interactions, we introduce band-symmetric ($+$) and band-antisymmetric ($-$) bosons~\cite{chou2024composite},
\begin{equation}\begin{split}
    \phi_\pm &= \tfrac{1}{2}\left[(\phi_{1\uparrow}\pm\phi_{2\uparrow})-(\phi_{1\downarrow}\pm\phi_{2\downarrow})\right],\\
    \vartheta_\pm &= \tfrac{1}{2}\left[(\phi_{1\uparrow}\pm\phi_{2\uparrow})+(\phi_{1\downarrow}\pm\phi_{2\downarrow})\right].
\end{split}\end{equation}
In this basis, the double spin-flip interaction is $\propto\cos(2\vartheta_-)$, and the interacting edge Hamiltonian density can be expressed in terms of two decoupled Luttinger liquids
\begin{equation}\begin{split}
    \mathcal{H}_{\mathrm{edge}} = &\sum_{s=\pm}\frac{v_s}{4\pi}\left[K_s(\partial_x\vartheta_s)^2+\frac{1}{K_s}(\partial_x\phi_s)^2\right] \\ &+ g \cos(2\vartheta_-).
    \label{eq:Lpm}
\end{split}\end{equation}
The velocities and Luttinger parameters are related to Eq.~\ref{eq:V} as
\begin{equation}\begin{split}
    v_\pm &= \sqrt{[(v+v')\pm(w-w')][(v-v')\pm(w+w')]},\\
    K_\pm &= \sqrt{\frac{(v-v')\pm(w-w')}{(v+v')\pm(w+w')}}.
\end{split}\label{eq:vg}\end{equation}
The double spin-flip cosine interaction has scaling dimension $2/K_-$. When the interaction is irrelevant ($K_-<1$), the $\phi_{\pm}$ and $\vartheta_\pm$ bosons all remain gapless, as do the fermion operators in Eq.~\ref{eq:Psi}. However, when the double spin-flip interaction is relevant ($K_->1$), the cosine term flows to strong coupling, such that $\vartheta_-$ is pinned and $\phi_-$ has short-range correlations~\cite{haldane1995stability}.  This is the bosonic edge phase. Since $\vartheta_-$ is neutral under both charge and $S^z$-spin conservation, the bosonic edge phase has the same symmetries as the original fermionic edge. However, for the bosonic edge phase, all single-fermion excitations are gapped. This can be confirmed by noting that the operators in Eq.~\ref{eq:Psi} all have short-range correlations in the bosonic phase, due to their dependence on $\phi_-$. While single fermions are gapped, the bosonic edge phase is still gapless in the band-symmetric sector spanned by $\phi_{+}$ and $\vartheta_{+}$. Notably, this implies that the following electron bilinears are still gapless:
\begin{equation}\begin{split}
    \psi^\dagger_{1\uparrow}\psi^\dagger_{2\uparrow} &\propto e^{-i(\phi_++\vartheta_+)},\\
    \psi^\dagger_{1\downarrow}\psi^\dagger_{2\downarrow} &\propto e^{i(\phi_+-\vartheta_+)},\\
    \psi^\dagger_{1\uparrow}\psi^\dagger_{1\downarrow} &\sim \psi^\dagger_{2\uparrow}\psi^\dagger_{2\downarrow} \propto e^{-i\vartheta_+},\\
    \psi^\dagger_{1\uparrow}\psi_{2\downarrow} &\sim \psi^\dagger_{2\uparrow}\psi_{1\downarrow} \propto e^{-i\phi_+}.
\end{split}\label{eq:pairs}\end{equation}
The first two operators are chiral bosons carrying charge $2e$ and spin $\pm1$, effectively spin-triplet Cooper pairs. The third and fourth operators are non-chiral bosons with charge $2e$ and spin $0$, and charge $0$ and spin $1$, respectively. These are a spin-singlet Cooper pair and a spin-flip, respectively. Since the lowest charge carried by a gapless local operator is $2e$, the bosonic edge phase is predicted to have Fano factor $F=2$ in quantum point contact shot noise~\cite{Tsantilas_2026,zhang2017fingerprints}.

\section{Segment geometry for edge numerics}\label{sec:segmentGeo}
Here we will present the setup used to numerically simulate the DQSHI edge using segment DMRG. While we will restrict our attention to the DQSHI, this setup can be directly generalized to other topological phases, as we discuss in Sec.~\ref{sec:discussion}.

\subsection{Cylinder Hamiltonian}\label{subsec:cylinderH}
To realize the DQSHI edge in a local and symmetric fashion, we consider the following Hamiltonian defined on an infinite 2D cylinder with periodic boundary conditions in the $y$-direction,
\begin{equation}\begin{split}
&H=H_0 + H_\mathrm{conf} + H_{\mathrm{int}}.
\end{split}\end{equation}
$H_0$ describes the DQSHI bulk, $H_\mathrm{conf}$ is a confining potential that creates an interface between the DQSHI and the vacuum, and $H_{\mathrm{int}}$ contains the two-body interactions needed for realizing the bosonic edge phase.

To realize the DQSHI bulk we use the following first-quantized Hamiltonian that is diagonal in the band and spin basis,
\begin{equation}
    H^{\tau \sigma}_0 = \frac{(\mbf{p}-\sigma_z \mbf{A})^2}{2m},
\end{equation}
where $\mbf{A} = ( 0,x/\ell^2)$. This is simply the Hamiltonian of a particle with mass $m$ in a spin-dependent magnetic field with magnetic length $\ell$. The DQSHI is realized upon filling all four degenerate LLLs. As depicted in Fig.~\ref{fig:segment_geo}(b), the confining potential $H_\mathrm{conf}$ is taken to be
\begin{equation}
    H^{\tau \sigma}_\mathrm{conf} =
\begin{cases}
  -\alpha L_x   & - L_{x} \geq x, \\
  \alpha x   & L_{x} \geq  x \geq -L_{x}\\
  \infty   & x \geq L_{x}
\end{cases},
\end{equation}
where the edge region $-L_x \leq x \leq L_x$ has width $2L_x$ in the $x$-direction, and, as we shall show, $\alpha$ sets the bare velocity of the edge modes.

Before introducing the interaction terms, $H_{\mathrm{int}}$, it will be useful to first project onto the low-energy subspace and switch to a second-quantized notation. Assuming that the potential varies slowly on the scale of the magnetic length, $\alpha\ell \ll \hbar\omega_c = \hbar^2/m\ell^2$, the low-energy subspace is spanned by the following LLL wave functions
\begin{equation}\begin{split}
    u^{\tau \sigma}_{m}(\mbf{r})&=\frac{e^{i k_{\sigma m} y}e^{-\left(x - \sigma_z \ell^2 k_{\sigma m} \right)^2/2\ell^2}}{\sqrt{L_y\sqrt{\pi}\ell}},\\ k_{\sigma m}&=\sigma_z \frac{2\pi \left( m +\frac{1}{2}\right)}{L_y},
\end{split}\label{eq:singleParticleWFs}\end{equation}
where $L_y$ is the circumference of the cylinder. In this definition, the $u^{\tau \sigma}_{m}$ wave functions all have the same guiding-center coordinate, $x_m = 2\pi (m + \frac{1}{2}) \ell^2/L_y$, but opposite spins have opposite momentum, $k_{\uparrow m} = -k_{\downarrow m}$.

Projecting onto the LLL wave functions, and switching to second-quantized notation, the single-particle Hamiltonian can be written as\footnote{For brevity, we have ignored the zero-point energy of the LLL in Eq.~\ref{eq:SecondQuantSinglePart}.}
\begin{equation}\begin{split}
H_0 + H_{\mathrm{conf}} &= \sum_{\tau \sigma}\sum_m E^{\tau \sigma}_m \,c^{\tau\sigma\dagger}_mc^{\tau \sigma}_m,\\
 \quad E^{\tau \sigma}_m &= \begin{cases}
  -\alpha L_x & - L_{x} \geq x_m, \\
  \alpha x_m  & L_{x} \geq  x_m \geq -L_{x}\\
  \infty   & x_m \geq L_{x}
  \end{cases},
\label{eq:SecondQuantSinglePart}\end{split}\end{equation}
where $c^{\tau \sigma}_m$ is the annihilation operator associated with the wave function in Eq.~\ref{eq:singleParticleWFs}. For states in the middle region, the energies can be rewritten as $E^{\tau \sigma}_{m} = \sigma_z v_{\mathrm{edge}} k_{\sigma m} $, with $v_{\mathrm{edge}} = \alpha \ell^2$, as expected for the helical edge modes. Due to the factor of $\ell^2/2L_y$ in $x_m$, half the edge states have negative energies while the other half have positive energies.

Let us assume generic interactions that, potentially, mix different $\sigma$- and $\tau$-sectors. Upon projecting onto the LLL wave functions, switching to second-quantized notation, and normal ordering, such interactions can be written as
\begin{equation}
H_{\mathrm{int}}=\sum_{\substack{\eta_1,\eta_2\\ \eta_3,\eta_4}}\sum_{\substack{m_1,m_2 \\ m_3,m_4}}V^{\eta_1 \eta_2; \eta_3 \eta_4}_{m_1,m_2;m_3,m_4}:  {c_{m_1}^{\eta_1\dagger}} c^{\eta_2}_{m_2}{c_{m_3}^{\eta_3\dagger}} c^{\eta_4}_{m_4}:,
\label{eq:SecondQuantInt}\end{equation}
where we have introduced the composite index $\eta = \tau\sigma$ to reduce notational clutter. Assuming that the bare interaction is short-ranged in real space (independent of the $\sigma$ and $\tau$ dependencies), the matrix elements can be written as
\begin{equation}\begin{split}
    &V^{\eta_1 \eta_2; \eta_3 \eta_4}_{m_1,m_2;m_3,m_4}\\ &=  U^{\eta_1 \eta_2; \eta_3 \eta_4}\int d\mbf{r}u^{\eta_1*}_{m_1}(\mbf{r}) u^{\eta_2}_{m_2}(\mbf{r}) u^{\eta_3*}_{m_3}(\mbf{r}) u^{\eta_4}_{m_4}(\mbf{r}),
\label{eq:ContactInt}\end{split}\end{equation}
where $U^{\eta_1 \eta_2; \eta_3 \eta_4}$ is the bare interaction strength. Following the bosonization analysis, we are interested in the double spin-flip interaction, Eq.~\ref{eq:dsf}, and density-density interactions, Eq.~\ref{eq:densitydensity}. The double spin-flip interaction corresponds to $\eta_1 = 1\uparrow$, $\eta_2 = 2\downarrow $, $\eta_3 = 1\downarrow $, and $\eta_4 = 2\uparrow$ (modulo trivial permutations). The density-density interactions correspond to $\eta_1 = \eta_2$ and $\eta_3=\eta_4$. In this setup, interactions are present everywhere throughout the cylinder. However, as we shall discuss in the next subsection, the effects of interactions can be ignored in the bulk, provided that the interaction scale is small compared to $\alpha L_x$.

In closing, we note that we have assumed that the $\tau = 1$ and $2$ bands are degenerate. Because of this, the Fermi momenta are the same for both $\tau$-sectors, $k_{F1\sigma} = k_{F2\sigma}$. This assumption is not important for our analysis. Carrying out the same construction, but with non-degenerate $\tau = 1$ and $2$ bands, leads to similar edge modes, except with $k_{F1\sigma} \neq k_{F2\sigma}$. This would cause the $\tau = 1$ and $\tau = 2$ modes to separate in space, weakening the interactions between them. However, for the short-range interactions in Eq.~\ref{eq:ContactInt}, the changes to the interaction strength are negligible, provided that $|k_{1\sigma} - k_{2\sigma}| \ll  \ell^{-1} $. Since $\ell$ is expected to be proportional to the lattice scale (moir\'e scale for twisted systems), $\ell^{-1}$ is a large momentum scale here. 

\begin{table}
    \centering
    \begin{tblr}{
      width = \linewidth,
      colspec = {c c c c},
      cells = {valign = m},
      rowsep = {3pt},
    }\toprule
       Excitation & $\Delta N_e$ & $\Delta S^z$ & {Bosonic edge phase} \\\midrule
       Single fermion & $1$ & $\frac{1}{2}$ &  gapped \\
       Singlet Cooper pair & $2$ & $0$ & gapless \\
       Spin flip & $0$ & $1$ & gapless  \\
       Triplet Cooper pair & $2$ & $1$ & gapless \\\bottomrule
    \end{tblr}
    \caption{\textbf{Gap structure for different types of edge excitations}. For each excitation type, the relevant gap is defined as in Eq.~\ref{eq:gapDef}, with a given $\Delta N_e$ and $\Delta S^z$. Due to time-reversal symmetry, $E_{\mathrm{gap}}(\Delta N_e, \Delta S^z) = E_{\mathrm{gap}}(\Delta N_e, -\Delta S^z)$. The last column indicates whether an excitation is expected to be gapped or gapless in the bosonic edge phase. All excitations are gapless in the fermionic edge phase.}
    \label{tab:gap_definitions}
\end{table}

\subsection{Segment DMRG}
We will now discuss how to use the second-quantized Hamiltonian defined in Eqs.~\ref{eq:SecondQuantSinglePart} and~\ref{eq:SecondQuantInt} to implement DMRG on the DQSHI edge. To begin, let us first ignore interactions, and initialize the 2D cylinder by filling all negative-energy single-particle states, i.e., those with $x_m < 0$. In this case, the low-energy degrees of freedom are the LLL edge modes. Next, we partition the cylinder into a semi-infinite DQSHI bulk segment, $ -L_x \geq x$, a semi-infinite vacuum segment, $x \geq L_x$, and a finite edge segment, $L_x \geq x \geq  -L_x$, such that the bulk segment is completely filled, the vacuum segment is completely empty, and the edge segment is partially filled. To analyze the effects of interactions, we then perform DMRG on the partially filled finite edge segment only. The bulk and vacuum segments are kept fixed---completely filled and empty, respectively---and serve as boundary conditions for the edge segment.

In this construction, we are assuming that interactions do not strongly affect the bulk or vacuum wave functions. This assumption is energetically justified when the interaction strengths $U^{\eta_1 \eta_2; \eta_3 \eta_4}$ are all small compared to $\alpha L_x$. This can be confirmed by noting that, in the absence of interactions, changing the bulk occupancy involves moving a particle from the bulk to the nearest unoccupied state, which will be near $x = 0$. This will incur an energy cost of $\alpha L_x$. We can therefore assume that the occupancy of the bulk is unchanged when the interaction strength is small compared to $\alpha L_x$. The validity of this assumption can be verified by showing that the local occupancy of the DMRG-optimized edge segment matches that of the bulk near $x = -L_x$. Similarly, the local occupancy should approach $0$ near the vacuum interface at $ x= L_x$.

Preempting our later discussion of the charge and spin gaps, we note that the total Hilbert space of the finite edge segment can be divided into sectors with well-defined total particle number, $N_e$, total $S^z$-spin, $S^z$, and momentum, $P$. We denote the ground-state energy in such a sector as $E_0( N_e,S^z,P)$. It will be useful to further define
\begin{equation}
    E_0(N_e,S^z )=\min_P E_0( N_e,S^z ,P),
\end{equation}
which is the ground-state energy, keeping only the total particle number and spin fixed.

The cost of the edge-segment calculation is set by the MPS bond dimension $\chi$ at the segment boundary, which is inherited from the bulk and vacuum segments, and by the entanglement generated within the edge segment. For the integer DQSHI considered here, the fixed bulk and vacuum segments are product states, so the cost is dominated by the edge segment, whose entanglement grows with $L_y$. In addition, the LLL-projected interaction yields a matrix product operator (MPO) with a bond dimension of order $100$, compared to order $10$ for Heisenberg or Hubbard models on cylinders, which further increases the cost (Appendix~\ref{app:numerics}).

\section{Results}\label{sec:results}
We will now numerically verify the existence of the bosonic edge phase of the DQSHI.
To do this, we consider the LLL-based setup of Sec.~\ref{sec:segmentGeo}, including the short-ranged interactions given in Eq.~\ref{eq:SecondQuantInt}, with
\begin{equation}
    U^{1\uparrow, 1\uparrow; 2\downarrow, 2\downarrow} = U^{2\uparrow, 2\uparrow; 1\downarrow, 1\downarrow} = U^{1\uparrow, 2\downarrow; 1\downarrow, 2\uparrow} = U_0,
\label{eq:intParam}\end{equation}
and all other $U$'s are set to zero. For this parametrization, there is only a single relevant dimensionless ratio, $U_0/\alpha \ell$. At weak coupling, $U_0/\alpha \ell \ll 1$, we can directly relate the microscopic model used here to the bosonized model in Sec.~\ref{sec:bosonicEdgeReview}: $v = v_{\mathrm{edge}} = \alpha \ell^2$, $w' = U_0\ell$, $g = U_0$ and $v' = w = 0$. Since the Luttinger parameter of the band-antisymmetric sector is $ K_- \approx 1 +  \frac{U_0 }{\alpha \ell}$, the double spin-flip interaction is relevant for $U_0 > 0$. In this case, $U^{1\uparrow, 1\uparrow; 2\downarrow, 2\downarrow}$ and $U^{2\uparrow, 2\uparrow; 1\downarrow, 1\downarrow}$ are repulsive density-density interactions.

To determine if the bosonic edge phase is realized in our DQSHI setup, we calculate the single-fermion and fermion-pair gaps of the edge using segment DMRG. Calculations are performed on the cylinder geometry with fixed $L_x = 12\ell$. As we are interested in thermodynamic gaps of the DQSHI edge, we consider a range of different edge lengths $L_y$ and extrapolate to the $L_y \rightarrow \infty$ limit, keeping the edge segment half-filled, $N_e = 4\lfloor L_x L_y/2\pi\ell^2 + 1/2 \rfloor$. We considered system sizes up to $L_y = 6\pi \ell \approx 18.8 \ell$ ($N_e=144$). All of the presented results were obtained with a bond dimension of $\chi = 4000$. Algorithmic details and the computational cost are given in Appendix~\ref{app:numerics}. Details of the extrapolation to the thermodynamic limit are given in Appendix~\ref{app:thermoLimit}.

Since the ground state is in the $S^z = 0$ sector for all values of $U_0/\alpha \ell$, the relevant gaps are
\begin{widetext}
\begin{equation}
    E_{\mathrm{gap}} (\Delta N_e, \Delta S^z ) = \frac{1}{2}\left( E_0 (N_e+ \Delta N_e, \Delta S^z) + E_0 (N_e- \Delta N_e, - \Delta S^z ) - 2 E_0 (N_e, 0 ) \right).
\label{eq:gapDef}\end{equation}\end{widetext}
Note that $E_{\mathrm{gap}} (\Delta N_e, \Delta S^z ) = E_{\mathrm{gap}} (\Delta N_e, -\Delta S^z )$ due to time-reversal symmetry. In the bosonic phase, $E_{\mathrm{gap}} (1, 1/2 )$ should extrapolate to a finite value in the thermodynamic limit since all single-fermion excitations are gapped. However, $E_{\mathrm{gap}} (2, 0 )$, $E_{\mathrm{gap}} (0, 1) $, and $ E_{\mathrm{gap}} (2, 1 )$ should all vanish due to the gapless excitations in Eq.~\ref{eq:pairs}. For the fermionic edge phase, all $E_{\mathrm{gap}}$'s should vanish in the thermodynamic limit. The gap structure of the fermionic and bosonic edge phases is summarized in Table~\ref{tab:gap_definitions}.

\begin{figure}
    \centering
    \includegraphics[width=1.0\linewidth]{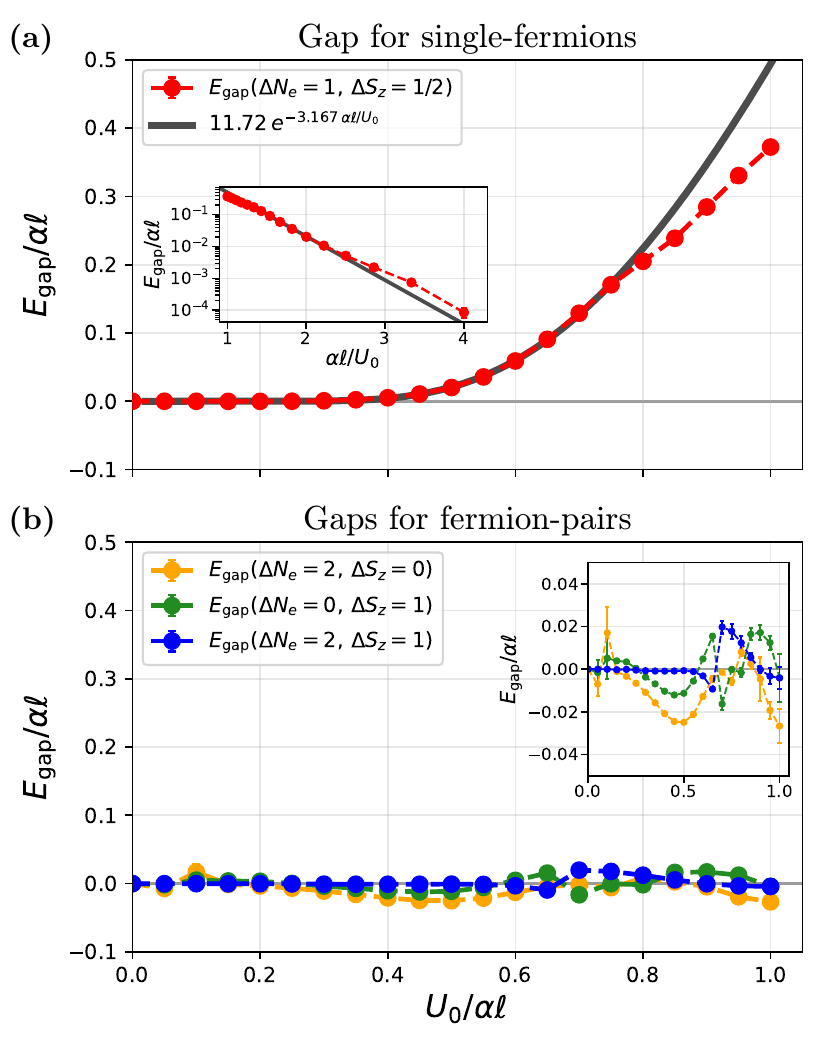}
    \caption{\textbf{Excitation gaps for the interacting DQSHI edge}. The gaps for single-fermion excitations \textbf{(a)} and fermion-pair excitations \textbf{(b)} as a function of the dimensionless interaction strength $U_0/\alpha \ell$, extrapolated to the thermodynamic limit. Error bars are obtained as the propagated errors from the least-squares fit. The solid line is a fit of $E_{\mathrm{gap}}(1,1/2)$ to $\Delta_0 \exp(-\kappa\,\alpha\ell/U_0)$ over $0 < U_0/\alpha\ell \leq 0.75$. The inset of  \textbf{(a)} shows $E_{\mathrm{gap}}(1,1/2)$ on a logarithmic scale as a function of $\alpha\ell/U_0$, where the exponential form appears as a straight line. The inset of \textbf{(b)}  magnifies the fermion-pair gaps, which scatter around zero within $0.04\,\alpha\ell$. Negative fermion-pair gaps are due to the $L_y \rightarrow \infty$ extrapolation, and all calculated gaps are positive for finite $L_y$ (see Appendix~\ref{app:thermoLimit}). }
    \label{fig:energies_repulsive}
\end{figure}

The extrapolated edge gaps are shown in Fig.~\ref{fig:energies_repulsive} for $0 \leq U_0/\alpha\ell \leq  1$. Calculations were performed with code based on {ITensors.jl}~\cite{Fishman_2022,ITensor-r0.3} as well as {TensorKit.jl}~\cite{Devos_2025} and {MPSKit.jl}~\cite{Devos_2026}.

As can be seen in Fig.~\ref{fig:energies_repulsive}, $E_{\mathrm{gap}}(1,1/2)$ increases with increasing $U_0/\alpha \ell$, while the fermion-pair gaps, $E_{\mathrm{gap}}(2,0)$, $E_{\mathrm{gap}}(0,1)$, and $E_{\mathrm{gap}}(2,1)$, all remain close to zero. This is exactly the behavior predicted for the bosonic edge phase of the DQSHI. The scaling of $E_{\mathrm{gap}}(1,1/2)$ with $U_0/\alpha \ell$ also matches theoretical predictions. Renormalization group analysis of the bosonized theory predicts a single-fermion gap $\propto \exp(-\pi \alpha \ell/U_0)$ for small $U_0/\alpha \ell$~\cite{fradkin2013field}. Fitting $E_{\mathrm{gap}}(1,1/2)$ to $\Delta_0 \exp(-\kappa\,\alpha\ell/U_0)$ over $0 < U_0/\alpha\ell \leq 0.75$, we indeed find $\Delta_0/\alpha\ell = 11.72 \pm 0.48$ and $\kappa = 3.167 \pm 0.028$. Including larger values of $U_0/\alpha \ell$ in this fit causes the root-mean-square error to quickly rise, signaling a departure from the weak-coupling regime.

\begin{figure}
    \centering
    \includegraphics[width=1.0\linewidth]{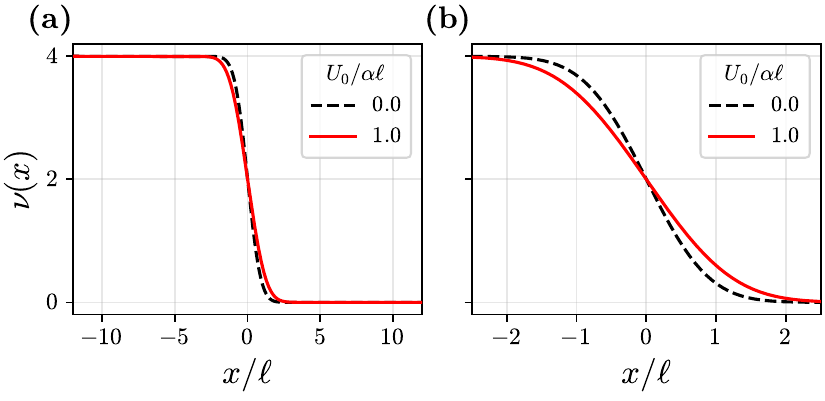}
    \caption{\textbf{Local density at the edge}. The spatially resolved electron density $\nu(x)$, summed over all four spin and band flavors, measured across the DQSHI edge for $L_y = 20 \pi \ell/3$ ($N_e = 160$) and interaction strengths $U_0/\alpha \ell = 0.0$ and $U_0/\alpha \ell = 1.0$, shown over the full edge region $(\bm{a})$, and zoomed in near $x = 0$ $(\bm{b})$. The density approaches the vacuum value of $\nu = 0$ as $x$ approaches $L_x = 12\ell$, and the bulk value of $\nu = 4$ as $x$ approaches $-L_x$.}
    \label{fig:density}
\end{figure}

To confirm that our use of segment DMRG is appropriate for the values of $U_0/\alpha\ell$ considered here, we plot the position-resolved electronic density for $U_0/\alpha\ell = 0.0$ and $1.0$ in Fig.~\ref{fig:density}. In both cases, the density quickly approaches $4$ (the density of the filled bulk states) as $x \rightarrow -L_x$. We can thus confirm that the edge wave function extrapolates to that of the bulk near the end of the edge segment. Similarly, the densities all quickly approach $0$ as $x \rightarrow L_x$, as expected for approaching the vacuum boundary. This justifies keeping the bulk and vacuum segments fixed even for the largest value of $U_0$ considered here.

\section{Discussion and outlook}\label{sec:discussion}
We have shown that, at an explicit microscopic model level, interactions can induce a bosonic edge phase in DQSHIs. In tailored segment DMRG simulations, we find the characteristic structure predicted by bosonization: gapped single fermions and gapless fermion pairs. The scaling of the single-fermion gap with interaction strength also matches theoretical predictions.
This numerical confirmation motivates the possibility that the bosonic edge phase can be realized in real DQSHIs, specifically in twisted MoTe$_2$ and WSe$_2$~\cite{Kang_2024_WSe2,Kang_2024_MoTe2,Kang_2025_TRSB,xu2025interplay}. We note that our explicit DQSHI model assumed that both sets of Chern bands were degenerate, unlike in the moiré TMDs. However, as pointed out in Sec.~\ref{subsec:cylinderH}, our results still hold if the bands are shifted in energy relative to each other.

Ultimately, whether or not the bosonic edge phase can be realized in the TMD-based DQSHIs will depend on how the different edge modes interact with each other. For single-QSHI edges in InAs/GaSb quantum wells~\cite{li2015observation, vayrynen2016magnetic} and bismuthene on SiC(0001)~\cite{stuhler2020tomonaga}, the extracted Luttinger parameters show strong repulsive interactions between the edge modes as a consequence of large bulk gaps and hence small edge localization lengths imposing dimensional quenching of Coulomb interactions. Such repulsive interactions can be beneficial to the formation of the bosonic edge phase: Repulsion between opposite spins in opposite bands stabilizes the bosonic edge phase, while repulsion between opposite spins in the same band instead can destabilize it. Determining which of these effects turns out to be dominant in the twisted TMD-based DQSHIs is an important future work task. 

If the interactions in these systems are suitable for stabilizing the bosonic edge phase, shot-noise
measurements should yield a Fano factor $F = 2$~\cite{zhang2017fingerprints, Tsantilas_2026}. Because
the doubling is an interaction effect, it is also tunable: weakening the interactions shrinks the
single-fermion gap, and once the gap falls below the bias voltage or temperature, the Fano factor
should cross over to $F = 1$. In practice, the interaction strength can be tuned by screening the
Coulomb interaction with a metallic gate~\cite{stepanov2020untying}. Observing a doubled Fano factor
that reverts to $F = 1$ under increased screening, or at elevated bias or temperature, would provide
strong evidence for the bosonic edge phase.

Finally, the segment construction is not specific to the DQSHI. For any topological phase whose bulk ground state is available as an infinite matrix product state (MPS), obtained from a microscopic Hamiltonian using iDMRG~\cite{McCulloch2008,Zaletel2013,Zaletel2015} or VUMPS~\cite{ZaunerStauber2018}, or constructed exactly for model wave functions~\cite{Zaletel2012,Estienne2013,Crepel2018,Crepel2019a,Crepel2019b}, the filled bulk segment can be replaced by this MPS, which then serves as the boundary condition for the edge segment~\cite{phien2012infinite, mong2017fibonacci, zhu2020topological, divic2025anyon}. This opens a route to studying anomalous edges of more general, including fractionalized, topological phases.

\begin{acknowledgments}
A.F.~and N.R.~acknowledge helpful discussions with M.~Stoudenmire and M.~Fishman. A.F.~thanks L.~Devos for his help in updating MPSKit.jl. This work is supported by the Deutsche Forschungsgemeinschaft (DFG, German Research Foundation) through Project-ID 258499086 -- SFB 1170 and through the W\"urzburg--Dresden Cluster of Excellence on Complexity, Topology and Dynamics in Quantum Matter -- ctd.qmat Project-ID 390858490 -- EXC 2147.
The Flatiron Institute is a division of the Simons Foundation.
Some of the work by A.F.~was conducted while at Google LLC. This research was conducted in the author's personal capacity; the views expressed are solely those of the author and do not represent the official policy or position of Google LLC or any other institution.
\end{acknowledgments}

\appendix

\section{Benchmark against ED}\label{app:sec:ED}

\begin{figure}
    \centering
    \includegraphics[width=1.0\linewidth]{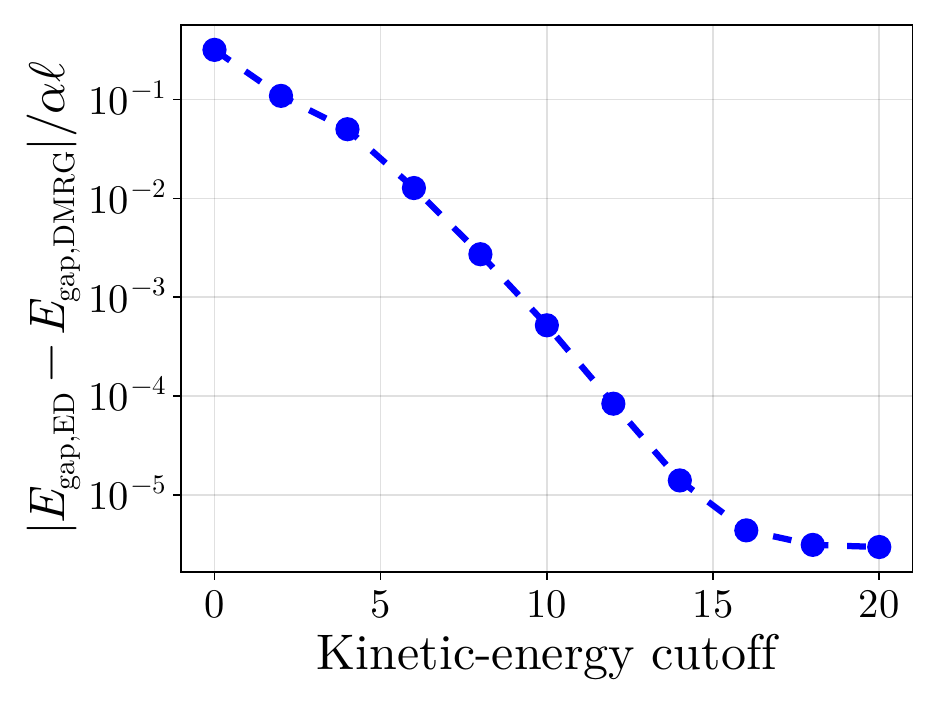}
    \caption{\textbf{Benchmark of segment DMRG against ED.} Absolute difference between the single-fermion gaps obtained with ED and segment DMRG as a function of the kinetic-energy cutoff of the ED. We consider $N_e = 64$ electrons in a half-filled segment at $U_0/\alpha\ell = 1$. For ED, the Hilbert space of the segment is truncated by a kinetic-energy cutoff, i.e., the maximum number of orbitals by which electrons can move away from the filled half of the segment. The Hilbert space dimension for the largest cutoff is $2.6\times 10^6$.}
    \label{fig:app:ED_benchmark}
\end{figure}

To validate the segment DMRG setup, we compare the single-fermion gap $E_{\mathrm{gap}}(1,1/2)$ obtained with DMRG against exact diagonalization (ED) on a truncated Hilbert space in Fig.~\ref{fig:app:ED_benchmark}. 

As the kinetic-energy cutoff of the ED is increased, the difference between the two gaps decays exponentially. The ED gap thus converges rapidly to the DMRG result, until it saturates at the DMRG truncation error. This calculation was performed on the smallest system size ($N_e=64$, $L_y\approx 8.4\ell$) used in the extrapolation in Appendix~\ref{app:thermoLimit}. 

\section{Extrapolation to the thermodynamic limit}\label{app:thermoLimit}

\begin{figure*}[th]
    \centering
    \includegraphics[width=1.0\linewidth]{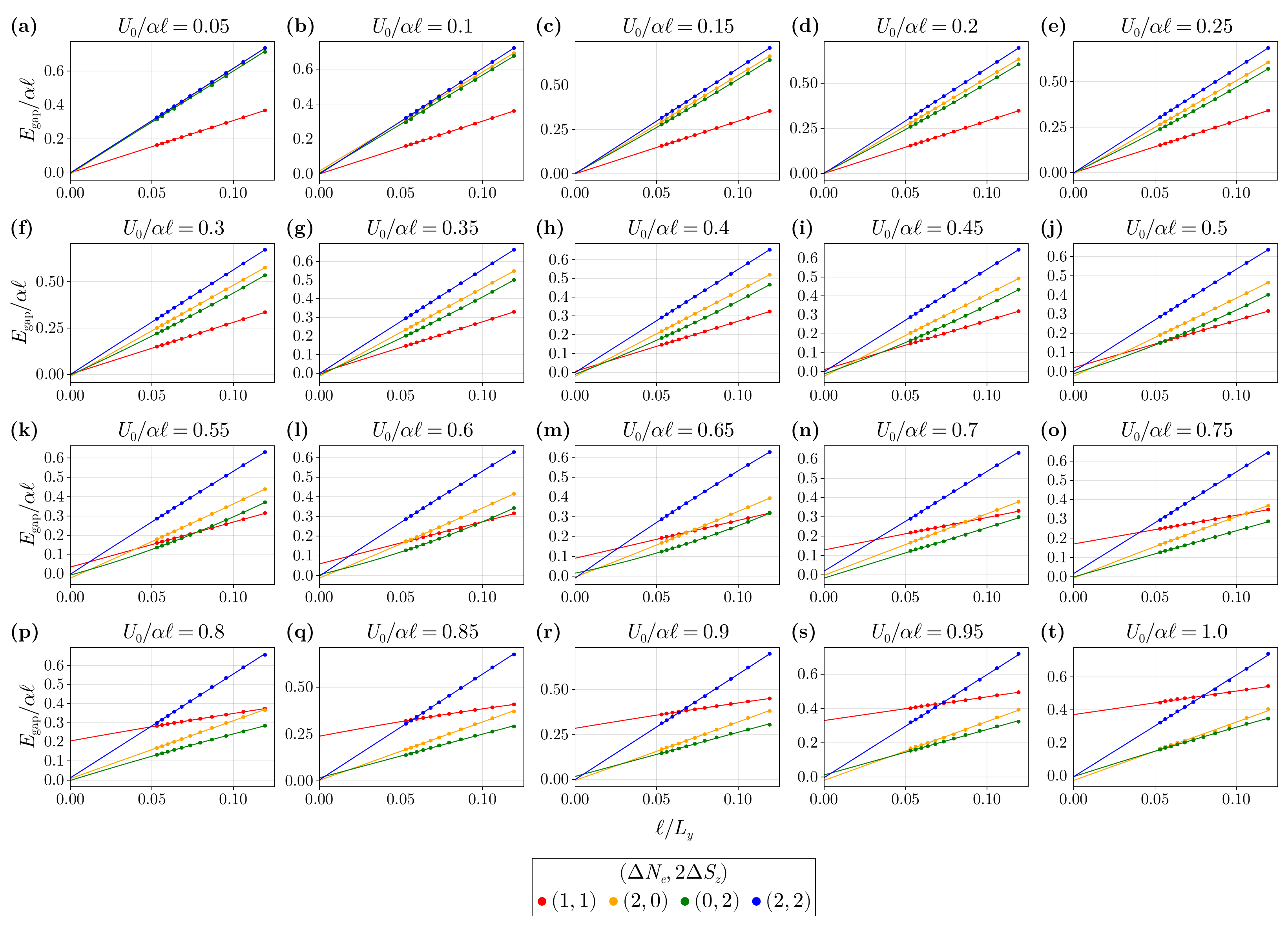}
    \caption{\textbf{Extrapolation of the edge gaps to the thermodynamic limit.} Gaps $\Delta E/\alpha\ell$ as a function of $1/L_y$ for (a)--(t) $U_0/\alpha\ell = 0.05$ to $1$ in steps of $0.05$. Colors denote the charge and spin sectors of the excitation, labeled by $(\Delta N_e, \Delta S^z)$. Markers show the segment DMRG data and lines the fits selected according to the protocol described in Appendix~\ref{app:thermoLimit}. We consider circumferences from $L_y\approx 8.4$ to $18.8\ell$, corresponding to $N_e=64$ to $144$ electrons.}
    \label{fig:app:repulsive_TD_limit}
\end{figure*}

Our segment DMRG calculations probe the physical edge, which forms a ring of circumference $L_y$, so we extrapolate the gaps in $1/L_y$. For each gap, we fit the data with both a linear and a quadratic polynomial in $1/L_y$,
\begin{align}
    f_\text{lin}(1/L_y)  &= \Delta_\infty^\text{(lin)} + \frac{a_1}{L_y}, \\
    f_\text{quad}(1/L_y) &= \Delta_\infty^\text{(quad)} + \frac{a_2}{L_y} + \frac{b_2}{L_y^2},
\end{align}
where $\Delta_\infty^\text{(lin)}$ and $\Delta_\infty^\text{(quad)}$ are the extrapolated gaps.

In most cases, the two fits agree. When they do not, we choose between them as follows. If one of the fits extrapolates to a clearly negative gap, below $-0.2\,U_0/\alpha\ell$, we take the fit whose extrapolated gap is closer to the gap at the largest perimeter we computed. Otherwise, we take the fit with the smaller extrapolated gap in magnitude. The second choice is deliberately conservative: when the fits disagree, the quadratic fit tends to follow finite-size oscillations and can overestimate the gap, so a finite gap that survives this choice is not an artifact of the fit.

For all parameters considered, the selected fits do not bend artificially as $1/L_y \to 0$. The finite-size gaps and the selected fits are shown in Fig.~\ref{fig:app:repulsive_TD_limit} for the parameters discussed in the main text.

\section{Computational details}\label{app:numerics}

The MPS bond dimension $\chi=4000$ understates the cost of our calculations, which is dominated by the size of the MPO. The LLL-projected contact interaction couples all four flavors and orbitals over a range that grows with $L_y$. Constructing the MPO exactly as a finite-state machine with merged equivalent paths~\cite{Crosswhite_2008a,Hubig_2017,Ren_2020} yields, for the largest calculation ($N_e=144$, $U_0/\alpha\ell=1$), a graph-compressed bond dimension $D_W=4039$. The SVD-based compression of Refs.~\onlinecite{Parker_2020,Soejima_2020} with cutoff $10^{-8}$ reduces this to $D_W=592$. For comparison, Heisenberg and Hubbard models on cylinders have $D_W$ of order $10$~\cite{Motruk_2016}. Since both the time and the memory of a DMRG step scale linearly with $D_W$ at leading order~\cite{schollwock2011density,Zaletel2015}, our calculations are more than an order of magnitude more expensive than for such models at equal $\chi$.

We use four techniques to reduce this cost. First, following Ref.~\onlinecite{Zaletel2015}, we split each orbital, with its $2^4$-dimensional local Hilbert space, into four MPS sites with a two-dimensional local Hilbert space each, ordered as $1\!\uparrow$, $2\!\uparrow$, $1\!\downarrow$, $2\!\downarrow$. This reduces the local Hilbert space of the two-site update from $16^2$ to $2^2$. Second, we store all environments that are not part of the current update on disk. Third, we perform the two-site update as in quantum-chemistry DMRG~\cite{White_1999,Chan_2002,Chan_2016}: we first contract the left (right) environment with the MPO tensor of the left (right) site and fuse the physical and virtual indices. The effective Hamiltonian is then a sum over the MPO bond of products of these enlarged-block operators, on which the Lanczos step acts. Fourth, we grow the bond dimension with the strictly single-site subspace expansion of Ref.~\onlinecite{Hubig_2015}, in the randomized-SVD formulation of Ref.~\onlinecite{McCulloch_2024}.

Each run increases $\chi$ by about $10\%$ per sweep up to $\chi=4000$ and then alternates between sets of five sweeps with and without subspace expansion, $65$ sweeps in total. For the largest system ($N_e=144$, $L_y=6\pi\ell$, $U_0/\alpha\ell=1$), one sweep at $\chi=4000$ takes $20$--$25$ minutes on $16$ physical cores of a dual-socket AMD EPYC 7742 node ($2.25$\,GHz base, up to $3.4$\,GHz boost). A full run yields a single ground-state energy in $10$--$12$ hours on $16$ cores, i.e., about $175$ core-hours, with a peak memory of $38$\,GB, excluding the environments on disk. The four gaps in Eq.~\ref{eq:gapDef} require eight such energies: $E_0(N_e,0)$, $E_0(N_e\pm1,\pm\tfrac12)$, $E_0(N_e\pm2,0)$, $E_0(N_e\pm2,\pm1)$, and $E_0(N_e,1)=E_0(N_e,-1)$. A single data point $(U_0,L_y)$ therefore costs about $1400$ core-hours.

\bibliography{refs.bib}

\end{document}